\documentclass[preprint,12pt]{aastex}

\begin{document}

\title{Discovery of a New, Polar-Orbiting Debris Stream in the Milky Way Stellar Halo}

\author{
Heidi Jo Newberg\altaffilmark{\ref{RPI}},
Brian Yanny\altaffilmark{\ref{FNAL}}, \&
Benjamin A. Willett\altaffilmark{\ref{RPI}}
}

\altaffiltext{1}{Dept. of Physics, Applied Physics and Astronomy, Rensselaer
Polytechnic Institute Troy, NY 12180\label{RPI}}

\altaffiltext{2}{Fermi National Accelerator Laboratory, P.O. Box 500, Batavia,
IL 60510\label{FNAL}}

\shortauthors{Newberg, Yanny}

\begin{abstract}

We show that there is a low metallicity tidal stream that 
runs along $l=143^\circ$ in the South Galactic Cap, about 34 kpc
from the Sun, discovered from SEGUE stellar velocities.  Since the 
most concentrated detections are in the Cetus constellation, and the 
orbital path is nearly polar, we name it the Cetus
Polar Stream (CPS).  Although it is spatially coincident with the
Sgr dwarf trailing tidal tail at $b=-70^\circ$, the metallicities
([Fe/H] = -2.1), ratio of blue straggler to blue horizontal branch stars, 
and velocities of the CPS stars differ from Sgr.  Some CPS stars
may contaminate previous samples of Sgr dwarf tidal debris.  The 
unusual globular cluster NGC 5824 is located along an orbit fit to the 
CPS, with the correct radial velocity.

\end{abstract}


\section{Introduction}

Over the past decade, spatial substructure from tidally disrupted satellites
has been discovered in the Milky Way's stellar spheroid 
\citep{netal02,betal06a,2008ApJ...673..864J,2008arXiv0811.3965G}.  However,
we expect that a more detailed accretion history of the Milky Way can be assembled
by including the kinematics of the stars \citep{2001AJ....122.1397H}, since the
kinematic signature of each tidally disrupted satellite remains long after
the spatial density has become such a small fraction of the stellar halo
density that the stream cannot be identified from spatial information alone.

Recently, \citet{ynetal09b} noticed 
a co-moving population of low metallicity blue horizontal branch stars (BHBs) 
with positions and velocities near, but not coincident with, the Sagittarius 
dwarf spheroidal trailing tidal stream in the South Galactic Cap, while
studying spectra of Milky Way stars from the Sloan Digital Sky Survey
(SDSS; York et al. 2000) and the Sloan Extension for Galactic Understanding
and Exploration (SEGUE; Yanny et al. 2009a).  The piece of this stream that
nearly intersects with the Sgr tidal stream is at $(l,b)=(140^\circ, -70^\circ)$ 
and at a distance of 34 kpc from the Sun, with a line-of-sight,
Galactic Standard of Rest velocity $v_{gsr}=-50$ km s$^{-1}$ and metallicity
[Fe/H] $\sim$ -2.0 (see figures 13 and 17 of Yanny et al. 2009b).  In this 
paper, we explore the extent and kinematics of this new stream.

\section{Observations and Data Analysis}

We first identify SDSS/SEGUE data release 7 (DR7; Abazajian et al. 2009) 
spectra that are likely to be associated with the
new stream.  Figure 1 shows a color-magnitude diagram of all stellar objects
in the South Galactic Cap with essentially zero proper motion and surface
gravities of giant stars.  Most of these stars are members of the stellar halo.
Circled observations have the velocity and metallicity we expect for the new
tidal stream.

SDSS/SEGUE data has very complex criteria for selecting the stars for spectroscopic
observation, so structure in the distribution of stars in Figure 1 is 
dominated by selection effects.  Since it is 
not possible to know the velocity of a star and it is difficult
to determine the metallicity of a star before the spectrum is obtained, the
selection is blind to these two quantities; therefore substructure can be identified
by looking for regions of Figure 1 in which the ratio of circled to uncircled points
is high.  The three boxes labeled blue horizontal branch (BHB; $-0.3 < (g-r)_0 < 0.2, 
0.8 < (u-g)_0  < 1.6, 17.7 < g_0 < 18.4$), red giant branch (RGB; $-12.75(g-r)_0 +25.62 < g_0 < -12.75(g-r)_0+27.12$, $16.8< g_0< 17.8$), and lower red giant branch 
(LRGB; $0.47 < (g-r)_0 < 0.53, 18.5 < g_0 < 19.7$) in Figure 1 have a relatively 
high fraction of stars
likely to be in the tidal stream, and comparison with M92 and M3 fiducials 
\citep{anetal08}, shifted to 34 kpc,
shows that they are also likely to be from the same stellar population.
From the BHB fiducials we extracted from the \citet{anetal08} data and 
distance moduli from \citet{h96}, we estimate the absolute magnitude
of the BHBs in the color range $-0.3<(g-r)_0<-0.2$, where most of the BHBs
lie, is $M_{g_0}=0.45$.

We will later show that this tidal debris stream follows fairly constant 
Galactic longitude, which will justify our current choice of plotting the 
velocities and apparent magnitudes of the stars as a function of Galactic 
latitude.  The upper panel of Figure 2 shows the velocities of stars in 
the three color-magnitude boxes in Figure 1.  The ones with lower 
metallicity are circled.  The solid outline identifies stars with 
velocities of the Sgr trailing tidal tail (compare with Law, Johnston \& 
Majewski 2005; Yanny et al. 2009b; $60^\circ < \Lambda_\odot < 140^\circ$).  
The dashed outline
shows velocities of stars in the new stream.  At higher Galactic latitude we 
relied primarily on the locus of low metallicity RGBs to select the 
velocities of stars in the new stream.  The new stream has lower metallicity 
than those of the Sagittarius trailing tidal tail, as demonstrated by the 
fraction of larger to smaller, point-like symbols within the upper
outlined region compared with the lower region with Sgr velocities.  


We explore the distance to the tidal stream in the lower panel of Figure 2,
which shows $g_0$ vs. $b$ for the stars in the upper plot that are likely
stream members, and photometrically selected BHBs in the region of the
newly identified Cetus Polar Stream (CPS).  
We find an approximately linear relationship between $g_0$ and Galactic 
latitude ($g_0 = -0.0162 b + 17.09$) in this portion of the
stream.  Distances were estimated and assuming $M_{g_0}=0.45$. 
Distance estimates are
tabulated in Table 1, with only statistical errors included.  Distance errors
may be systematically too high or too low by 10\%, depending on the determination of the absolute magnitude of BHBs \citep{setal04}.

The four panels of Figure 3 show (upper left) an estimate of the positions of 
the F turnoff stars in the CPS, and the positions of the photometrically
selected BHB stars; (upper right) the $(l,b)$ distribution of spectra with colors
and magnitudes similar to those in the CPS; (lower left) the 
distribution of F turnoff stars in Sgr and the CPS, with the stars with
CPS velocities superimposed; and (lower right) the same F turnoff
stars with the stars with Sgr stream velocities superimposed.
Note that there is an overdensity of photometrically selected BHB stars
that lines up with the background-subtracted F turnoff star overdensity,
and the CPS velocity-selected BHB, RGB, and LRGB stars, running
approximately along Galactic latitude $l \sim 143^\circ$.  Stars that are
velocity selected to be candidate Sgr stream stars follow a different path
in the sky, along the Sgr dwarf tidal tail as tabulated in \citet{netal03}.
Although the two streams cross near $b=-70^\circ$, they are about $30^\circ$
apart at $b=-30^\circ$.  The lack of significant numbers of colored points 
at $l < 110^\circ$ or at $l> 160^\circ$ (where the Sagittarius stream 
is located) gives us confidence that this stream is not an artifact, 
and is distinct from the previously identified Sagittarius trailing tail.
The Galactic longitude of the stream center in each of the four SDSS stripes
76, 79, 82, and 86, was estimated by comparing
the positions of the F turnoff stars, photometrically selected BHB stars,
and velocity-selected BHB, RGB, and LRGB stars.
One sigma errors were also estimated by eye.  
The stream centers and estimated errors are given in Table 1.

To estimate the metallicity of the CPS we histogramed
spectroscopically selected candidate stream stars from Figure 3 that 
have $105^\circ < l < 160^\circ$.  The peak of the distribution for
BHB, RGB, and LRGB stars is [Fe/H] $\sim -2.1$.  The formal mean metallicity 
of the  stars is [Fe/H] $\sim -1.98 \pm 0.04$.  There is a population
of LRGB stars that is higher metallicity ([Fe/H] $\sim -1.3$).  If this
population is removed, then the formal mean is [Fe/H] $= -2.08\pm0.04$.
Figure 17 of \citet{ynetal09b} shows that the BHB stars in the CPS
are slightly more metal poor than those of the Sgr dwarf
trailing tidal stream, but both tidal streams may plausibly have a broad
distribution of stellar ages and metallicities.


Table 1 summarizes the properties of the CPS at four
Galactic latitudes, shown in Figures 2 and 3.
In addition to the position, velocity, and distance of the stream as
estimated from Figures 2 and 3, we list the approximate velocity 
dispersion of the line-of-sight velocities ($\sigma_v$), and the number of spectra 
at each location.
The velocity dispersion for each stripe 76, 79, 82, and 86 is  computed from 
the spectra with $120^\circ<l<165^\circ$ that are shown in the lower left
panel of Figure 3, and tabulated in Table 1.  Since the intrinsic SDSS/SEGUE radial 
velocity errors are about 4 km s$^{-1}$, the intrinsic velocity dispersion of the CPS
is about 4.5 km s$^{-1}$ in stripes 76, 82, and 86, and about 10 km s$^{-1}$
in stripe 79.

\section{Discussion}

The Cetus Polar Stream solves a puzzle long pondered by the authors.  In
\citet{ynetal00} we discovered the Sgr dwarf tidal tails along the
Celestial Equator, including the trailing tidal tail in stripe 82.  We
have always wondered why, in Fig. 3 of that paper, the BHB stars at
$g_0=18$ in the South Galactic Cap appear offset in position in the sky
from the Sgr blue straggler (BS) stars that are two magnitudes fainter.  The counts
of BHB and BS stars, as defined by \citet{ynetal00}, along southern
SDSS stripes 79, 82, and 86 are presented in Figure 4.  From this figure, we
determine that many of the SDSS BHB stars in the southern stripes that
had previously been attributed to the Sgr trailing tidal tail are
actually in the CPS.
The ratio of BS (higher surface gravity A-colored stars) 
to BHB stars varies amongst globular clusters 
(i.e. see Figures 12-16 of An et al. 2008), and can be used as an 
identifying marker in the halo to help separate two populations with
distinct origins or evolutionary histories.
From Figure 4, it is clear
that the Sgr trailing tidal tail has a much larger BS/BHB ratio than
the CPS.

We fit an orbit to the four CPS locations
in Table 1, following the procedure used by \citet{wetal09} and a 
spherical halo potential ($q=1.0$).  The average of the best fit
orbit from five random starts of the fitting algorithm is shown by
the solid black lines in Figures 2 and 3.
The formal chi-squared per degree of freedom for this orbit is 1.08.  Varying 
the halo flattening from 0.3 to 1.25 does not significantly change the 
goodness-of-fit, and changes the best fit orbit by less than the formal
errors.

The CPS stars in the lower left panel of Figure 3 are spread over quite a
range of Galactic latitudes (at least $15^\circ$, or $\sim 10$ kpc), 
which argues in favor of a dwarf galaxy progenitor, though the low velocity
dispersion ($\sim 5$ km s$^{-1}$) argues for a diminutive dwarf galaxy
or possibly a globular cluster.  No known dwarf galaxies lie close to the
best fit orbit, but the globular cluster NGC 5824, at
$(l,b) = (332.5^\circ,22^\circ)$ is located within $3^\circ$
of the orbit, at a very plausibly correct distance, and has a radial velocity
that matches the predicted orbit radial velocity within one sigma.  NGC
5824 has a well-populated BHB and a measured [Fe/H]=-1.85.  NGC
5824 measurements are taken from \citet{h96}.
Additionally, the tidal distortion of NGC 5824 measured by 
\citet{gfiq95} and \citet{lmc00} is aligned with the CPS orbit.
\cite{gfiq95} show that NGC 5824 has a 
central cusp; this massive globular cluster could have once been a dwarf 
galaxy core \citep{nGC}.  Alternatively, it could be associated
with the dwarf galaxy progenitor or be the sole progenitor.

\section{Conclusions}

A previously unknown, low metallicity tidal debris stream is identified
at $l=143^\circ$ and 34 kpc from the Sun in the South Galactic Cap.  Although 
it is spatially coincident with the 
Sgr dwarf trailing tidal tail at $b=-70^\circ$, the metallicities
([Fe/H] = -2.1), ratio of BS/BHB stars, and velocities of the Cetus Polar 
Stream stars are significantly different.  Some BHB stars that have been attributed
to the Sgr trailing tidal tail by previous authors are instead part of the
CPS.  Both the width of the tidal stream ($\sim 10$ kpc) suggests a
dwarf galaxy progenitor, though the
velocity dispersion ($\sigma \sim 5$ km s$^{-1}$) opens the possibility
for a globular cluster progenitor.  
The globular cluster NGC 5824 is located on the CPS orbit with the
correct radial velocity, distance, and plausible metallicity.  It is
additionally elongated along the orbit.  NGC 5824 could be the progenitor,
the core of a dwarf galaxy progenitor, or associated with a dwarf galaxy
progenitor.  This stream was
discovered from a study of SDSS/SEGUE velocities, which allowed us to 
separate it from the Sgr trailing tidal tail even though they intersect
spatially.

\acknowledgments 

This work was supported by the National Science Foundation, grant AST
06-06618.  We acknowledge several important suggestions from the anonymous referee.  This paper utilized data from the SDSS and SDSS-II databases 
(http://www.sdss.org).


\begin{deluxetable}{rrrrrrrr}
\tabletypesize{\scriptsize} \tablecolumns{8} \footnotesize
\tablecaption{Cetus Polar Stream Detections} \tablewidth{0pt}
\tablehead{
\colhead{$l$} & \colhead{$\delta l$} & \colhead{$b$} & \colhead{$\rm v_{gsr}$} &  \colhead{$\sigma_v$}  & \colhead{N} & \colhead{d}& \colhead{$\delta d$}  \\
\colhead{$^\circ$} & \colhead{$^\circ$} & \colhead{$^\circ$} & \colhead{$\rm km~s^{-1}$} & \colhead{$\rm km~s^{-1}$}  &\colhead{}&\colhead{kpc}& \colhead{kpc} 
}
\startdata
144&2&-71&-29.8&6.4&4&36.1&1.9\\
144&3&-62&-42.7&6.5&8&33.8&1.8\\
142&3&-54&-39.2&11.1&16&31.8&1.7\\
142&4&-46&-59.2&5.8&14&30.1&1.6\\
\enddata
\end{deluxetable}

\begin{figure}
\includegraphics[scale=0.75,angle=-90]{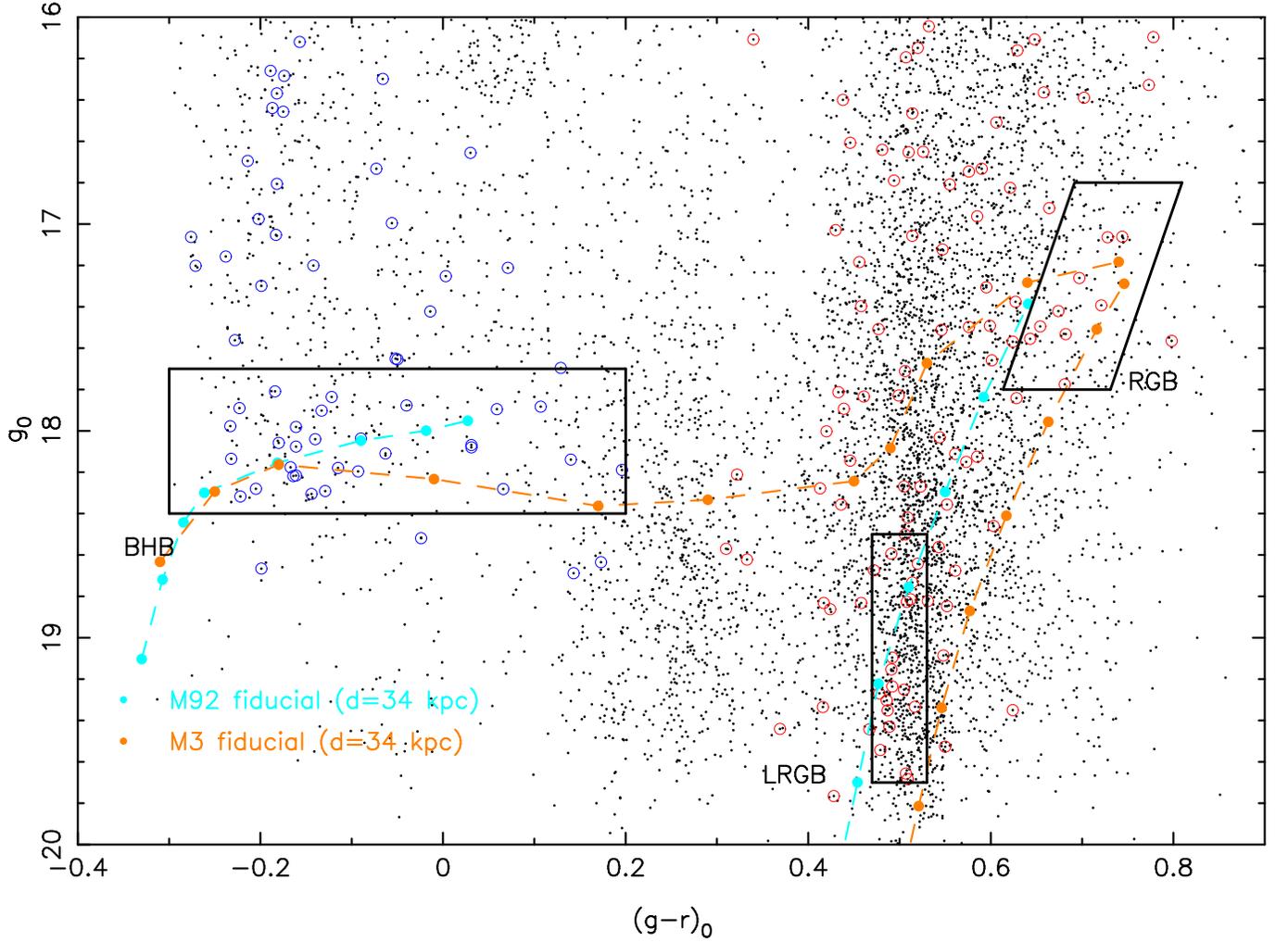}
\caption[Cetus Polar Stream candidate CMD]{
\footnotesize
We show as small black dots the color and $g_0$ magnitude for all the SDSS DR7 stellar
spectra in the South Galactic Cap ($b<0^\circ$) with surface gravities of giant stars
($1< \rm log~g < 4.0$), and essentially zero proper motion ($|\mu_l|<6$ mas yr$^{-1}$,
$|\mu_b|<6$ mas yr$^{-1}$).  These cuts select objects likely to be in the stellar
halo.  The circles show those points that have velocities and metallicities
consistent with membership in the new stellar stream ($-77<v_{gsr}<0$ km s$^{-1}$,
$\rm-4<~[Fe/H]~<-1.9$.  Stars with $-0.3 < (g-r)_0 < 0.2$ are likely BHB stars, so
for these stars we used the SDSS [Fe/H]$_{WBG}$ metallicity measurement (blue
circles).  The stars
with $0.3<(g-r)_0<0.8$ are likely giant stars, so in this color range we used the SDSS
[Fe/H]$_a$ metallicity measurement (red circles). 
Fiducial sequences for M92 and M3, shifted to 34 kpc, are shown for reference.
}
\end{figure}

\begin{figure}
\begin{center}
\includegraphics[scale=0.6,viewport=1in 0in 7in 7.5in]{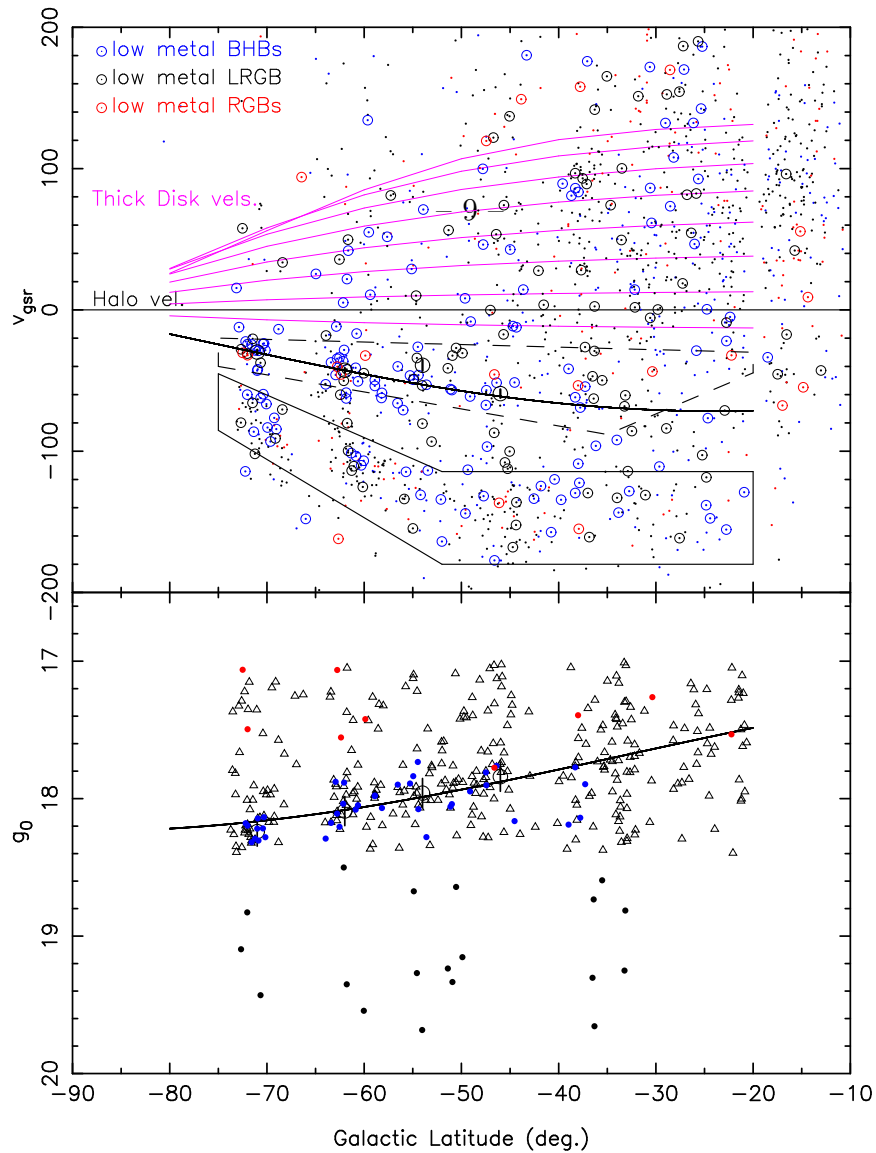}
\end{center}
\caption[Velocities and Distances to the new stream] {
\footnotesize
The top panel shows velocities of the stars in the color-magnitude boxes in Figure 
1. 
Spectra with metallicities of $\rm -4<[Fe/H]<-1.9 $
are circled.  (Note that the circled points are not selected in velocity as 
they were in Figure 1.)  The solid line boxed region shows stars with the velocities 
of the Sgr dwarf trailing tidal tail.  Note that many of the stars in this 
stream have metallicities higher than -1.9, and are therefore not circled.  
The boxed region with the dashed line shows the velocity of the newly identified tidal 
debris.  Note in particular that there 
are very few RGB candidates with low metallicity outside of the Sgr
stream and the new stream.  There are a few
low metallicity RGB stars with $v_{gsr} \sim 150$ km s$^{-1}$ that may have a
kinematic association with another (unrelated) structure.  We overlay the calculated velocities of thick disk stars 
with $50^\circ < l < 190^\circ$ (top to bottom) in magenta to show there is no confusion 
with stream candidates even if $\rm log~g$ is misidentified.  Filled circles in the 
lower panel show the apparent magnitudes of the stars
in the upper panel that have metallicities and velocities of the new stream
(all circled stars inside the dashed box in the upper panel, using the same color code).  
The triangles show photometrically selected BHB stars (see Yanny et al.
2000 for photometric selection technique) that have $120^\circ<l<165^\circ$.
The trend of distance with $b$ is consistent between photometrically
selected BHBs and those with spectra.
Note that many of the LRGB stars in the new stream are too faint to be included
in the SDSS/SEGUE spectroscopic survey, and the intrinsic magnitude distribution
of RGB and LRGB stars is very broad.  The adopted mean stream velocities (upper
panel) and apparent magnitude of the BHB stars (lower panel) are shown as open 
circles with errors, and
a solid black curve shows the best fit orbit through those stream locations in
both panels.
}
\end{figure}


\begin{figure}
\includegraphics[scale=0.7,viewport=1in 0.0in 7in 8.0in]{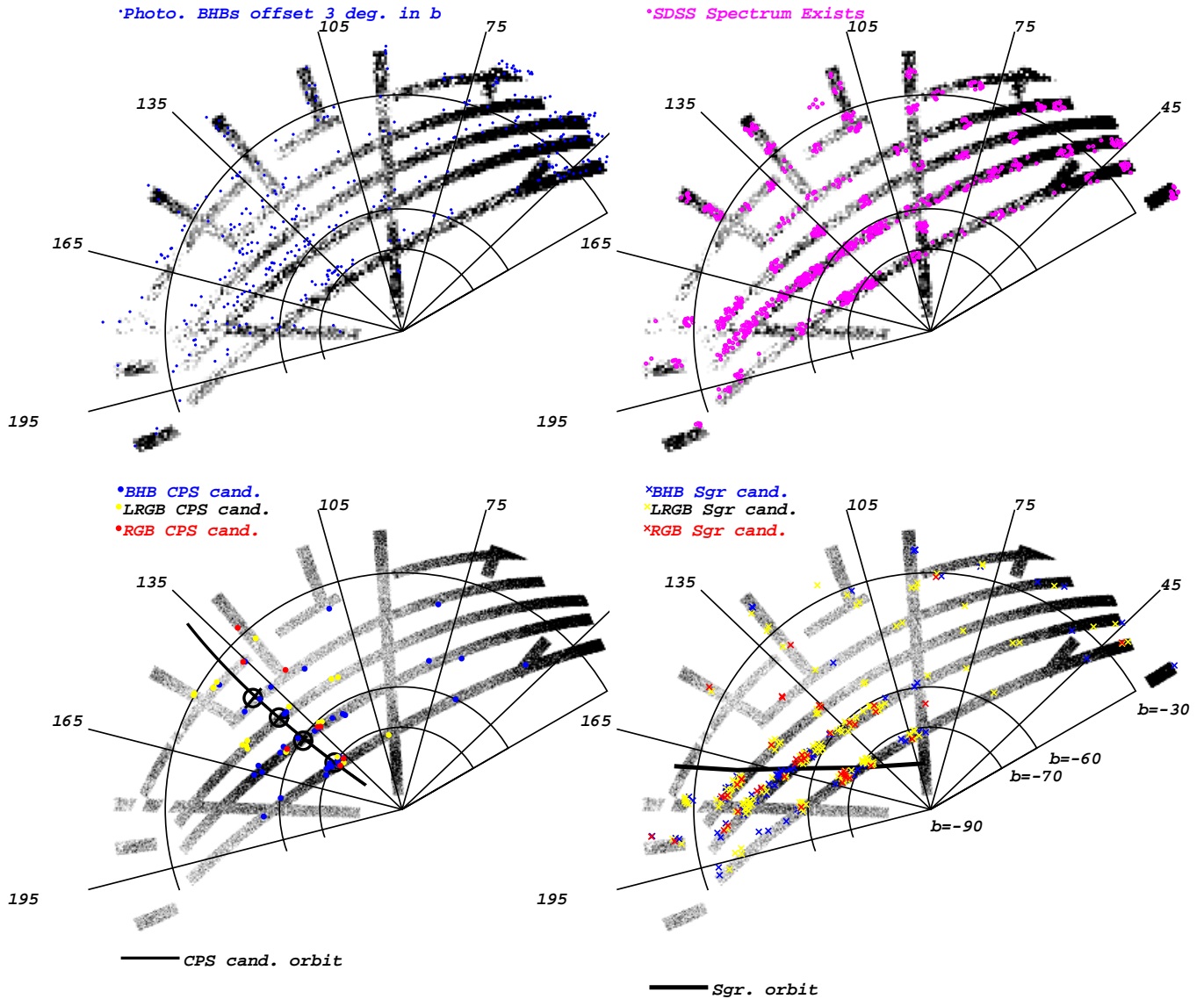}
\caption[Galactic coordinates of new stream] {
\footnotesize
The upper left panel shows the 
density of turnoff stars in a CPS color magnitude box 
($20.25 < g_0 < 21.5, 0.22 < (g-r)_0 < 0.36$) minus the density of 
stars in a Sgr box ($20.35 < g_0 < 21.85, 0.10 < (g-r)_0 < 0.20$)
in polar Galactic coordinates, origin at the SGP.
The overdensities (dark areas) running along $l=140^\circ$ from
$b=-70^\circ$ up to $b=-40^\circ$ show the CPS.
The blue dots in the upper left panel show the positions 
(offset $3^\circ$ in $b$ for clarity) of photometrically selected CPS candidate
BHBs, with $-0.0162 b + 16.94 < g_0 < -0.0162 b + 17.24$. Note the excess along 
$130^\circ < l < 150^\circ, -70^\circ < b < -40^\circ$.
The upper right panel shows (magenta circles) the locations
of stars with SDSS/SEGUE spectra in the color-magnitude selection boxes 
of Figure 1, showing the completeness coverage
of the spectroscopy relative to the imaging.  
The lower left panel shows the density of turnoff stars 
with $20.5 < g_0 < 22.5, 0.26 < (g-r)_0 < 0.30, (u-g)_0 > 0.4$, 
highlighting both the Sgr and Cetus debris streams (the Sgr
tidal stream is more prominent).
The filled circles show the stars
with velocities and metallicities consistent with membership in the new CPS,
color coded by spectral type.
The heavy black curve shows the best fit orbit for the CPS structure.  The 
lower right panel shows (crosses) the positions of low 
metallicity stars with spectra in the upper panel of Fig. 2 that have 
the velocities of the Sgr trailing tidal tail (the low metallicity
subset of SDSS/SEGUE Sgr spectra), along with a Sgr locus 
(heavy line). 
}
\end{figure}

\begin{figure}
\includegraphics[scale=0.6,angle=-90]{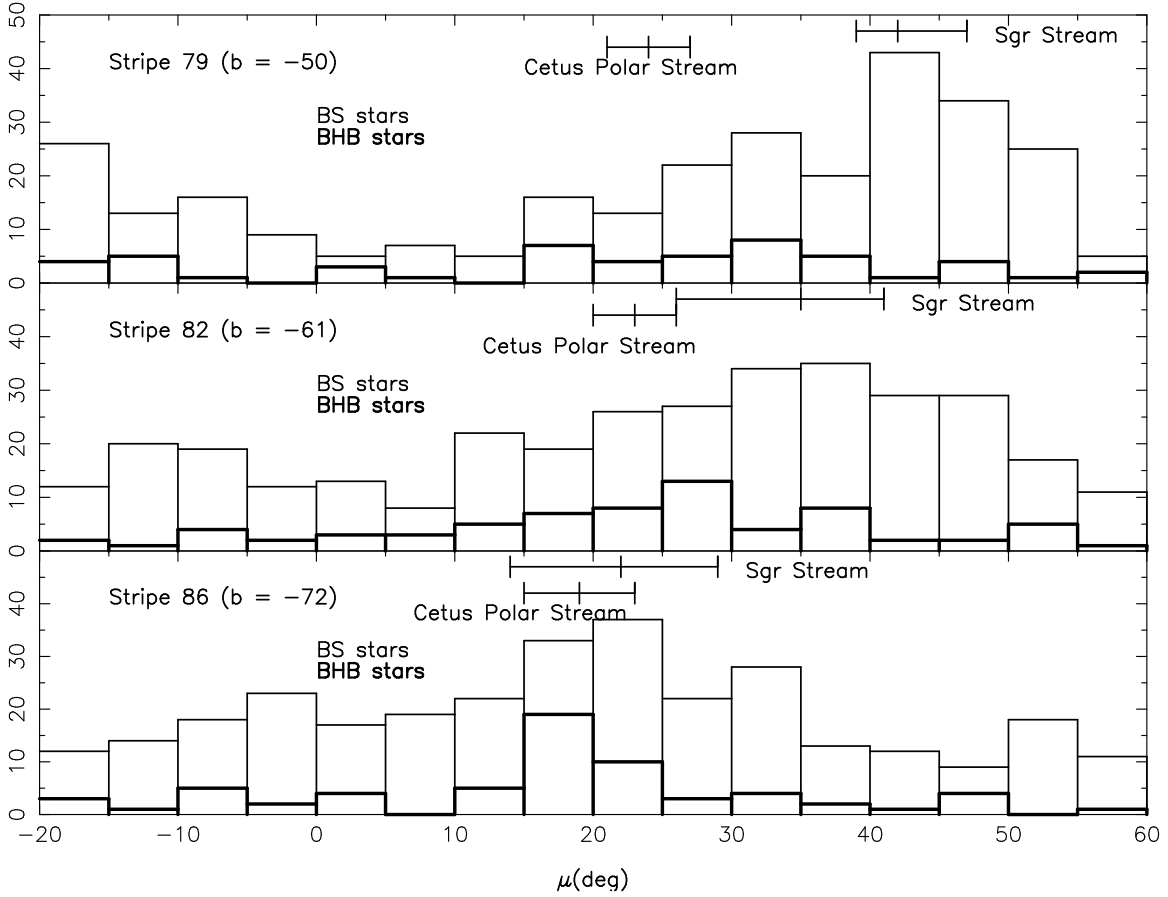}
\caption[BHB and BS stars in Sgr and Cet] {
\footnotesize
We show the number of photometrically selected BHB and BS stars in SDSS 
stripes 79, 82, and 86, in the South Galactic Cap.
The stars are selected from SDSS imaging data with 
$-0.3 < (g-r)_0 < 0.0,~ 0.8 < (u-g)_0 < 1.6$ and color separated into 
BHBs and BSs using the line in Figure 10 of \citet{ynetal00}, and the BSs are 2 mags fainter intrinsically than the BHBs.  The x-axis 
gives angular distance along the SDSS stripe (SDSS survey longitude, $\mu$),
which is identical to right ascension $\alpha$ for stripe 82, and deviates 
from $\alpha$ by less than two degrees for the nearby stripes 79 and 86.  
The regions where the Cetus Polar Stream and the Sgr stream 
intersect each stripe are indicated by the horizontal bars near the top of each 
panel.  Note that there is considerable overlap between the two streams 
in stripe 86 (lower panel), while they are largely disjoint in stripe 
79  (upper panel).  The relative number of BS/BHB stars, a ratio which 
may be regarded as an intrinsic property of a region of an individual 
stream, differs significantly between the two streams.
}
\end{figure}

\end{document}